\documentclass[12pt,preprint]{aastex}

\def\dfplot#1{\plotone{#1}}
\def\BE{\begin{equation}}
\def\BEL#1{\begin{equation}\label{#1}}
\def\EE{\end{equation}}
\newcommand{\COBE}{{\it COBE}}
\newcommand{\MAP}{{\it MAP}}

\newcommand{\HeI}{He\,{\scriptsize I}}
\newcommand{\HeII}{He\,{\scriptsize II}}
\newcommand{\HII}{H\,{\scriptsize II}}

\newcommand{\Halpha}{H$\alpha$}
\newcommand{\etal}{{\it et al.}~}

\newcommand{\degree}{^\circ}

\newcommand{\kms}{{\rm ~km/s}}
\newcommand{\MAG}{{\rm ~mag}}

\newcommand{\emunit}{{\rm ~cm^{-6}~pc}}

\newcommand{\JypSr}{{\rm ~Jy~sr^{-1}}}

\newcommand{\GHz}{{\rm ~GHz}}

\newcommand{\K}{{\rm ~K}}

\newcommand{\microK}{\mu{\rm K}}

\begin{document}

\title{A Full-Sky H-alpha Template for Microwave Foreground Prediction}

\author{Douglas P. Finkbeiner\footnote{Hubble Fellow}}
\affil{Princeton University, Department of Astrophysics,
Peyton Hall, Princeton, NJ 08544}


\begin{abstract}
A full-sky \Halpha\ map with $6'$ (FWHM) resolution is presented.
This map is a composite of the Virginia Tech Spectral line Survey
(VTSS) in the north and the Southern H-Alpha Sky Survey Atlas (SHASSA)
in the south.  The Wisconsin H-Alpha Mapper (WHAM) survey provides a
stable zero-point over 3/4 of the sky on a $1\degree$ scale.  This
composite map can be used to provide limits on thermal bremsstrahlung
(free-free emission) from ionized gas known to contaminate
microwave-background data.  The map is available on the WWW for public
use.

\emph{Subject headings: }
dust, extinction --- 
ISM: clouds ---
HII regions
\end{abstract}

\section{INTRODUCTION}
\label{sec_intro}
Over the last 5 years, 3 surveys of the \Halpha\ line (the 3-2
transition in neutral atomic H -- used as a tracer of ionized gas)
have revolutionized our knowledge of the warm ionized medium.
Two high resolution surveys reveal supernova remnants, supershells,
and filamentary structure in the diffuse ISM with breathtaking
detail.  A third survey, lower in spatial resolution but with velocity
information, has provided a dramatic picture of the kinematics of the
ISM over 3/4 of the sky. 

To the microwave astronomer, these surveys can play another role:
tracing the free-free emission of the Galaxy.  Cosmic microwave
background (CMB) anisotropy experiments have long detected ISM
emission as a foreground at $\sim10-50\GHz$
(\COBE, Kogut \etal\ 1996; Saskatoon, de Oliveira-Costa \etal\ 1997;
OVRO, Leitch \etal\ 1997; 19GHz survey, de Oliveira-Costa \etal\ 1998; 
Tenerife, de Oliveira-Costa \etal\ 1999).  In some cases this ISM-correlated
emission has been presumed to be free-free or synchrotron emission
\cite{kogut96}; in other cases, electric dipole emission
\cite{draine98} from rapidly rotating dust grains is suspected
\cite{doc99,finkbeiner02}.  With the increasing capabilities of
current and future microwave experiments (CBI, Mason \etal\ 2002;
DASI, Halverson \etal\ 2002; \MAP, Bennett \etal\ 1997) 
a more thorough understanding of Galactic foreground emission will
become increasingly important for cosmological work.

Now that the \Halpha\ data are public we have combined these 3 surveys
into a moderate resolution ($6'$) full-sky well-calibrated composite
map, optimized for use as a free-free template.  Version 1.1 of this
map, along with software to access it, is available on the World Wide
Web.\footnote{http://skymaps.info}

\section{THE DATA}
The full-sky composite \Halpha\ map is comprised of three wide-angle
surveys, two at high resolution with poorly determined zero-point
calibrations, and one at low resolution with a stable zero-point. 
The data used here were downloaded in December, 2001 from the web sites of
VTSS\footnote{http://www.phys.vt.edu/$^\sim$halpha}, 
SHASSA\footnote{http://amundsen.swarthmore.edu/SHASSA}, and
WHAM\footnote{http://www.astro.wisc.edu/wham}.

\subsection{VTSS}
The Virginia Tech Spectral line Survey (VTSS) is a survey of the
northern hemisphere with the Spectral Line Imaging Camera (SLIC) which
is a fast $(f/1.2)$ CCD camera with a narrow bandpass (17\AA) \Halpha\
filter as well as a continuum filter.  The CCD has a quantum
efficiency of 80\% at 6500\AA.  Survey images have $1.6'$ pixels in
$385\times 385$ images with a usable radius of 5 degrees from the
center of each pointing.  The fast optics and low noise CCD result in
sub-Rayleigh sensitivity at confusion limited levels \cite{vtss}. 
At present, 107 fields have been released on the VTSS web 
site.  As more are released, we will incorporate them into future
versions of this map. 

\subsection{SHASSA}
The complimentary effort in the south is the Southern H-Alpha Sky
Survey (SHASSA; Gaustad \etal\ 2001), using a small camera at Cerro Tololo
Inter-American Observatory (CTIO).  The survey consists of 542 fields
of $13\degree\times 13\degree$ spaced every $10\degree$ on each pass.
The two passes are shifted by $5\degree$ in each coordinate, covering
the southern $\delta < 20\degree$ sky twice.  Each image is approximately
1k$\times$1k with $47''$
pixels.  The sensitivity per pixel is $\sim2$ Rayleigh, similar to VTSS.  
The $5\degree$
offset between the passes allows one to discard the corners
of the images, which are more difficult to calibrate.  The SHASSA
survey is complete, and is available on the
web.

\subsection{WHAM}
The Wisconsin H-Alpha Mapper (WHAM) Northern Sky
Survey (Reynolds, Haffner, \& Madsen 2002) consists
of 37,565 spectra obtained with a dual etalon Fabry-Perot spectrometer
on a 0.6m telescope at Kitt Peak.  The velocity resolution of $12\kms$
makes possible the removal of the geocoronal \Halpha\ emission,
providing a stable zero-point unavailable in the other two surveys.
Velocity coverage is approximately $-100 < v_{LSR} < 80$ km/s and
includes all significant \Halpha\ emission along most lines of sight. 
The survey pointings are on a grid with $1\degree$ spacing,
undersampling the instrument's $1\degree$ diameter tophat beam (Haffner,
Reynolds, \& Tufte 2003).  


\section{PROCESSING STEPS}

Essentially the same steps are followed for VTSS and SHASSA, except
for the PSF (point spread function) wing masking for SHASSA.  In this
section we describe the treatment of stellar artifacts and the zero-point
calibration using WHAM, compare resulting data in the VTSS / SHASSA
overlap area, define an error map, and describe the reprojection of
the surveys to a full sky Cartesian grid.

\subsection{Stellar Artifacts}

\begin{figure}[tb]
\epsscale{0.9}
\dfplot{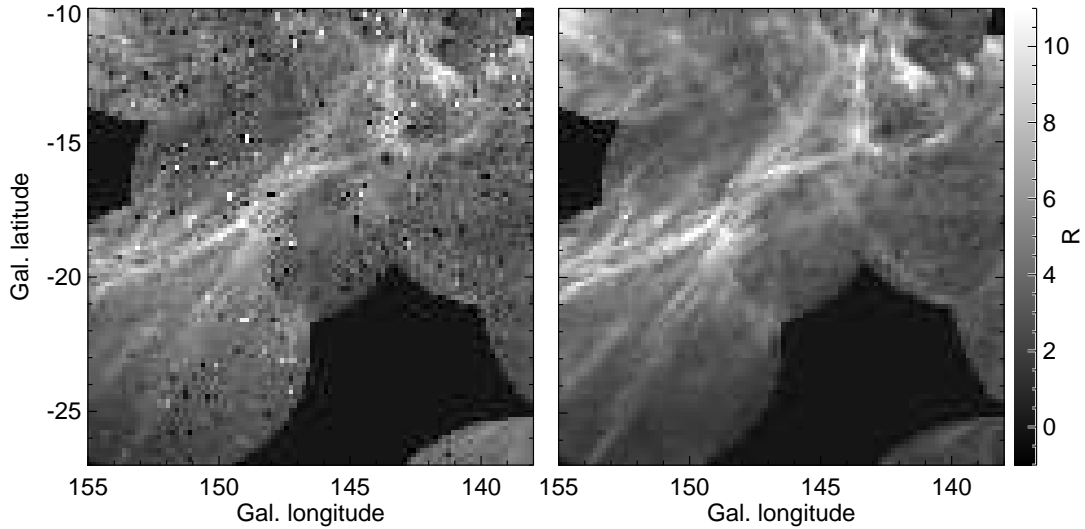}
\figcaption{VTSS images zero-pointed to WHAM without \emph{(left)} and
with \emph{(right)} point source removal.
\label{fig_v1}
}
\end{figure}

\begin{figure}[tb]
\epsscale{0.9}
\dfplot{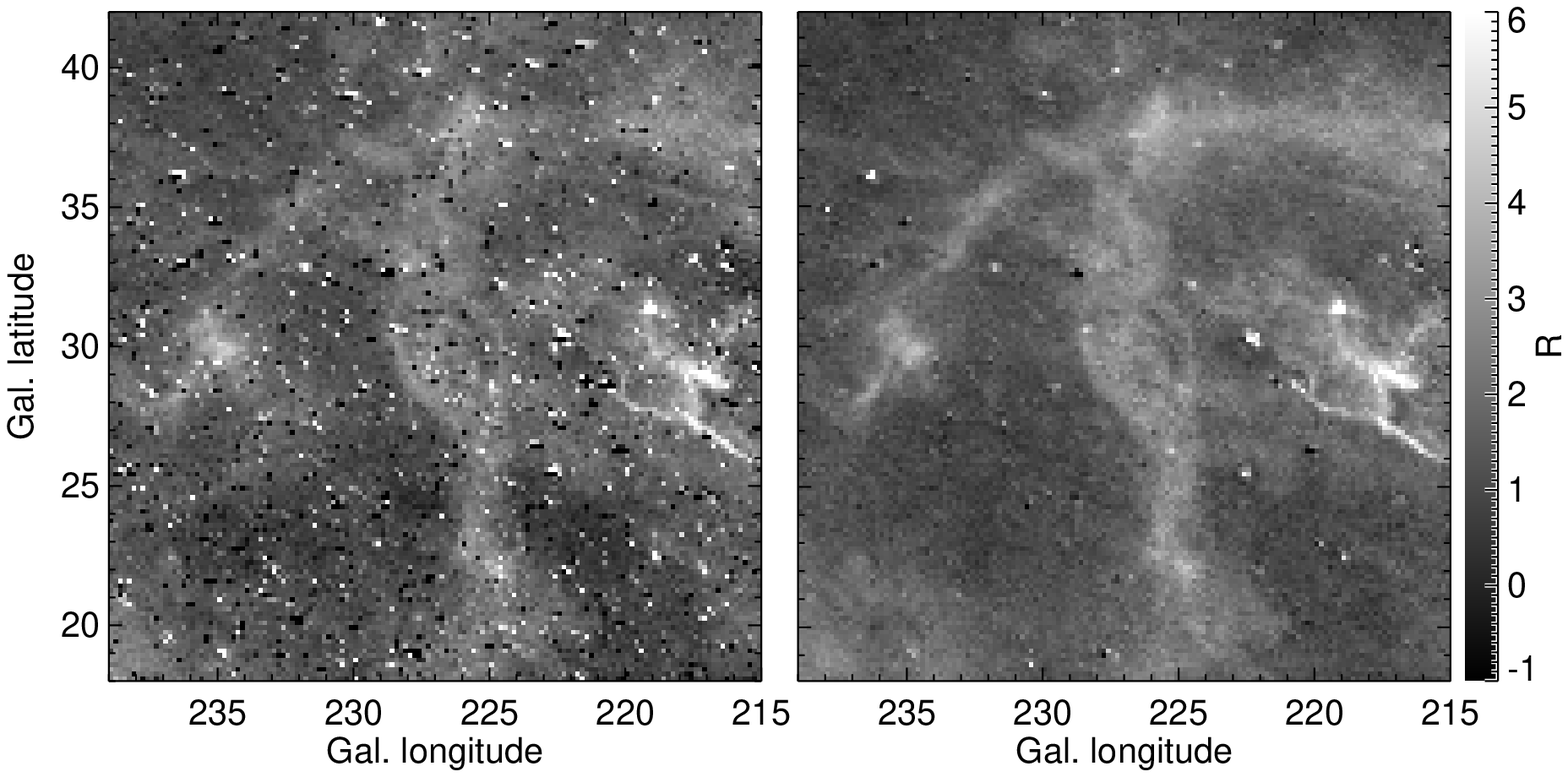}
\figcaption{SHASSA images zero-pointed to WHAM without \emph{(left)} and
with \emph{(right)} point source removal.  All
remaining stellar artifacts have the BRT\_OBJ bit set in the bitmask. 
\label{fig_s1}
}
\end{figure}

The VTSS and SHASSA surveys have already produced
``continuum-subtracted'' images, meaning that a properly scaled
continuum image has been subtracted from the narrow-band \Halpha\
image to remove stars.  Because of variations in PSF and stellar
color, there are positive and negative residual ghosts in the
subtracted image for $V \la 11$ stars.  These are small compared to
the original stellar
contamination (typically $10-20\%$) but large compared to some of the
ISM structure.  Removing these without damaging the real structure is
our challenge.

The poorly subtracted stars are difficult to locate in the subtracted
image, but are readily found in the continuum image with the
well-known DAOfind algorithm created by Peter Stetson, as implemented
in IDL by Wayne Landsman\footnote{In the Goddard IDL User's Library at
http://idlastro.gsfc.nasa.gov/}.  The continuum
images provided with the SHASSA survey are not generally co-registered with
the \Halpha\ images, and VTSS provides no continuum images at all, so
for source detection we simply used as a continuum image the
difference between the (already) co-registered \Halpha\ images and
``continuum-corrected'' \Halpha\ images.
The sources identified in these images are then removed from
the \emph{continuum-corrected} image by linear interpolation to the unmasked
parts of an annulus around them, with $3\sigma$ outlier rejection.
This allows the fit to follow the gradient of any underlying ISM
structure.  The radius of the replaced region is 2 pixels (about
$1.5'$ for SHASSA, $3'$ for VTSS), expanded to 3.5 pixels for Tycho
$V \la 9\MAG$.  For bright stars a region of radius $16'$, $20'$, and $30'$
is flagged in the BRT\_OBJ mask bit for star brighter than Tycho V of
5.5, 3, and $1.5\MAG$ respectively.  Because many of these bright stars
reside within complicated \HII\ regions, it would be reckless to
interpolate over such large areas of the map.  However, the remaining
residuals from stellar subtraction mean that these regions should be
excluded from statistical studies.  The BRT\_OBJ bit is only set in the
SHASSA area; the outer wings of the VTSS PSF are better behaved so no such
mask is necessary. 
Visual inspection of several fields indicates that the algorithm
works well and finds stars while ignoring real ISM structure.  Figure
\ref{fig_v1} shows VTSS data with and without the point source
subtraction; Figure \ref{fig_s1} displays the same comparison for
SHASSA.


\begin{figure}[tb]
\epsscale{0.6}
\dfplot{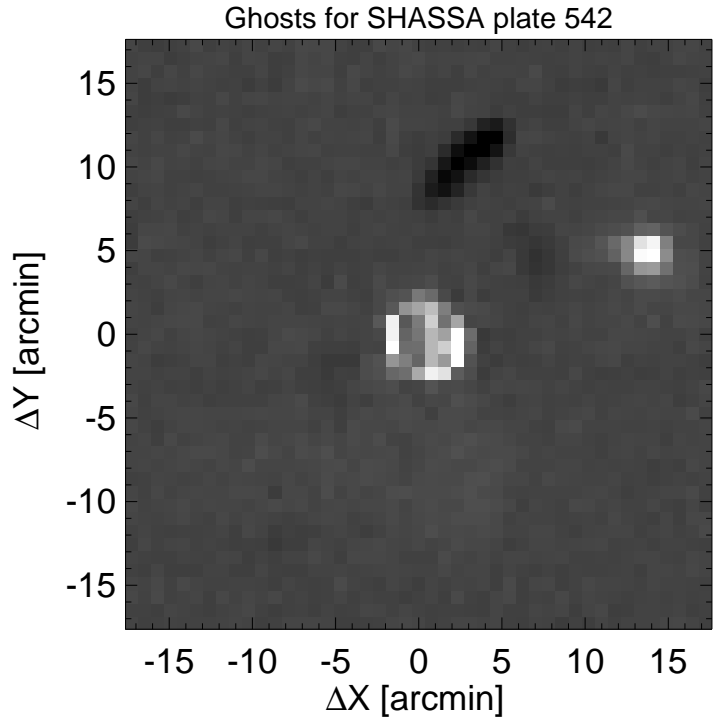}
\figcaption{
SHASSA continuum-subtracted PSF wing pattern for plate 542.  
The positive ghost is caused by an internal
reflection in the \Halpha\ filter.  The negative ghost is a reflection
in the continuum filter.  This pattern is typical, and is fairly
constant throughout the survey, but we determine it independently for
each SHASSA field.  In some fields it is rotated by 180 degrees. 
\label{fig_wing}
}
\end{figure}

In the case of the SHASSA survey, the wings of the PSF are not well
behaved.  Imperfections in the filters cause ``transmissive'' ghosts
near bright objects (Gaustad \etal\ 2001, \S2.4).  The ghosts in the
narrow-band and continuum images are displaced, causing a defocused
positive and negative ghost to appear in the wings of the
continuum-subtracted PSF.  The PSF is determined independently for
each image, and found to be quite stable (Figure \ref{fig_wing}).
Rather than attempting to deconvolve this PSF over the whole image, we
simply determine which pixels are deviant and mask those near known
objects.  A cut of Tycho $V < 7\MAG$ for stars was chosen by
inspection, and the same mask (determined plate by plate) is then
applied to all bright stars on a given plate. 

The 100 brightest galaxies in the sky can also be seen in the map, but the
flux from these is usually comparable to other remaining artifacts, so
they are not removed.  Many bright compact sources remaining in the
maps are not artifacts at all, but planetary nebulae, for example NGC
246 at $(l,b) = (118.86, -74.71)$ and NGC 7293 (the Helix Nebula; see
Speck \etal\ 2002) at
$(36.14, -57.15)$.  Such objects can be very bright (500R or more) but
pains have been taken to not remove them from the map.  Statistical
comparisons with microwave data sets may want to mask these out and
handle them separately.

Saturated pixels in bright stars require special handling.  Bleed
trails are present in both VTSS and SHASSA, but because of the
preprocessing done upstream from the publicly released data,
identification of saturated pixels is not trivial.  Saturation is
identified by taking the union of all pixels above a threshold, and
those pixels with a second-difference in the direction orthogonal to
the bleed trails above a threshold.  This mask is then extended for 2
pixels in all directions to deal with nearby contamination.  The
neighboring pixels in directions perpendicular to the bleed direction
are used for interpolation across the
masked pixels.  In most cases this yields acceptable results, even in
the presence of real structure.  The brightest few stars in the SHASSA
region have bleed trails of different length in each of 5 exposures, 
so it is not always possible to remove
these perfectly from the combined data.  A few sections of bleed
trails remain in the map around the brightest stars.  A full
re-analysis of the raw data would provide the required information,
and may be attempted if these few remaining artifacts are found to
compromise the scientific content of the present map. 

In only a few places (e.g. Orion) the \Halpha\ image itself has bleed
trails.  These regions must be handled on a case by case basis, and
masked to indicate no processing should be done.  No processing was
done on the LMC, SMC, or M31 for similar reasons.  These pixels have
the ``big\_obj'' mask bit set.

\subsection{The WHAM zero point}
\label{sec_whamzero}

Broadband surveys such as SHASSA and VTSS have an uncertain zero point
because of smooth foreground contamination from geocoronal \Halpha\
emission.  The spectral resolution of WHAM allows removal of this
foreground emission which is often brighter than the signal of
interest.  Note however that the WHAM data cover velocities $-100 <
v_{LSR} < 80$ km/s and there is significant \Halpha\ emission at
$v_{LSR} > 80$ km/s over a small part of the sky.  We have estimated an
upper bound on the error thus introduced by assuming that the
intensity in the 80 km/s velocity bin is constant for $80 < v_{LSR} <
120$ km/s.  Any WHAM pointing for which this upper bound exceeds both
0.2R and 10\% has bit 7 (HI\_VEL) set in the bit mask.  Apart from
this 0.5\% of the sky, WHAM can provide the correct zero-point for
high resolution images on one degree scales.  

Because the
WHAM survey is undersampled ($1\degree$ diameter beams on a $1\degree$ spacing)
it does not contain enough information for an unambiguous solution,
but the following prescription is reasonable, and produces an
aesthetically pleasing map consistent with reality.

We approximate the WHAM beam to be a
smoothed tophat with the form
\BE
W(r) = \frac{1}{\exp[(r-r_0)/r_s]+1}
\EE
where $r$ is the angular distance from the beam center, $r_0$ is $1/2$
degree, and $r_s$ is a smoothing term set to $1/40$ degree.  The
smoothing is desirable because it makes the WHAM beam fully sampled in
the raw \Halpha\ images during convolution. 

Each VTSS image is convolved with the WHAM beam centered on each WHAM
pointing.  The WHAM value is then subtracted (a floor of $-0.1$R is
imposed on a few spurious WHAM measurements), yielding a set of
differences, one for each WHAM pointing.  Each WHAM pointing is then
assigned a Gaussian (FWHM = $1\degree$) weight used to interpolate the
zero-point difference, which is then subtracted from the VTSS image.  
Again, this solution is not unique,
nor strictly correct.  A bandlimited WHAM-like data set could be
interpolated exactly with sinc interpolation, leaving no ambiguity,
but no such data exist at this time.  The SHASSA images are processed
in the same way, with the additional constraint that the corners of
the images ($r > 500$ pixels), which are systematically biased, are
smoothly de-weighted. 

Some outlier rejection is required in the above algorithm.  Very
bright sources can have a large difference residual, resulting in dark
halos in the final map.  Such
differences, if they deviate by more than 3 R from the median of
neighboring differences, are set to that median.  This is justified
because it removes unphysical halos and the error thus introduced in
the bright pixels is fractionally very small.

\subsection{Reprojection}

Full sky maps, masks, and weight arrays are generated for the three
surveys at a resolution of $6'$.  This smoothing is acceptable for a
CMB free-free template because $6'$ is higher resolution than current
or planned low-frequency CMB data sets (even Planck HFI will only be
$5'$ FWHM) and is convenient for comparisons with the $6'$ FWHM 
Schlegel, Finkbeiner, \& Davis (1998; hereafter SFD98)
dust map.

We use a full sky Cartesian Galactic $(l,b)$ projection with
$8640\times4320$ pixels ($2.5'$ square pixels at $b=0$). 
Each of the 542 SHASSA and 107 VTSS source images is gaussian smoothed
prior to projection, so that it will be well sampled ($2.4'$ per FWHM)
when reprojected.  Then the pixel centers from the destination image
are transformed to $(x,y)$ positions on the source image, where a
value is obtained for each via bilinear interpolation.  Because the
smoothed source image is heavily oversampled, the form of
interpolation is not important.  An apodized weight map for each
source image is also used so that subtle differences in the
zero-levels in each image are blended smoothly and do not create sharp
boundaries in the final image.

The WHAM survey itself must also be reprojected, so as to fill in
regions of the sky where no SHASSA or VTSS data are available.  A
Delaunay triangulation is performed on the 37565 WHAM pointings and
the IDL trigrid function \cite{renka83} is used to interpolate values
for every pixel.  Naturally, the interpolation over the southern sky
($\delta \la 30\degree$) is undesirable and must be masked.  We extend
the WHAM pointing grid to cover the southern area, set these ``mock''
pointings to zero weight, and interpolate with the same Delaunay
triangulation used for the data.  Any pixels with weight less than
unity after interpolation are masked out.

\begin{figure}[tb]
\epsscale{1.0}
\dfplot{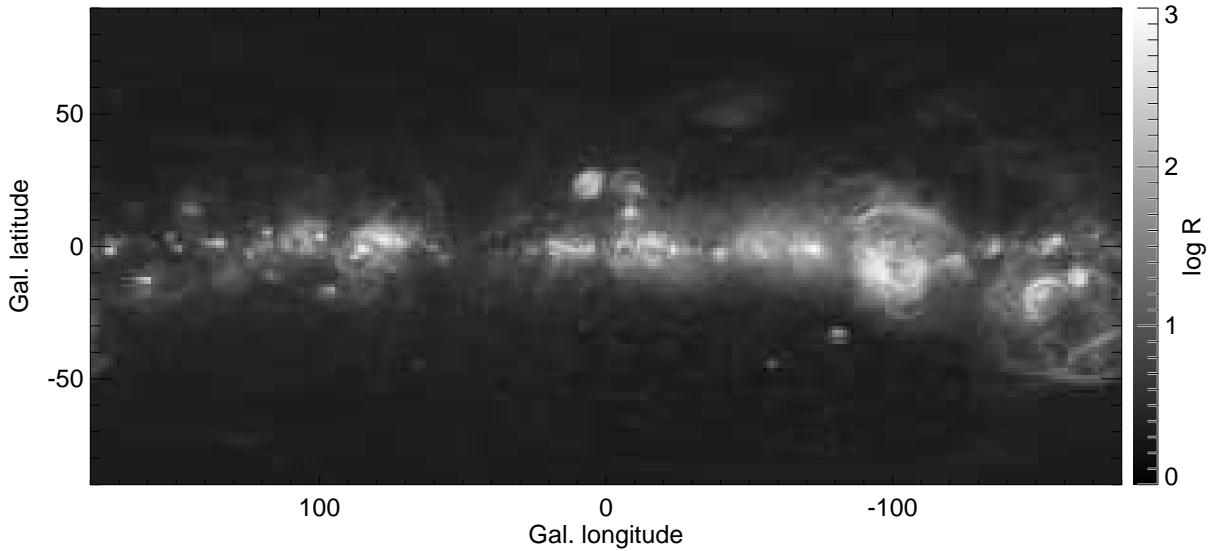}
\figcaption{The composite \Halpha\ map.
\label{fig_composite}
}
\end{figure}

\begin{figure}[tb]
\dfplot{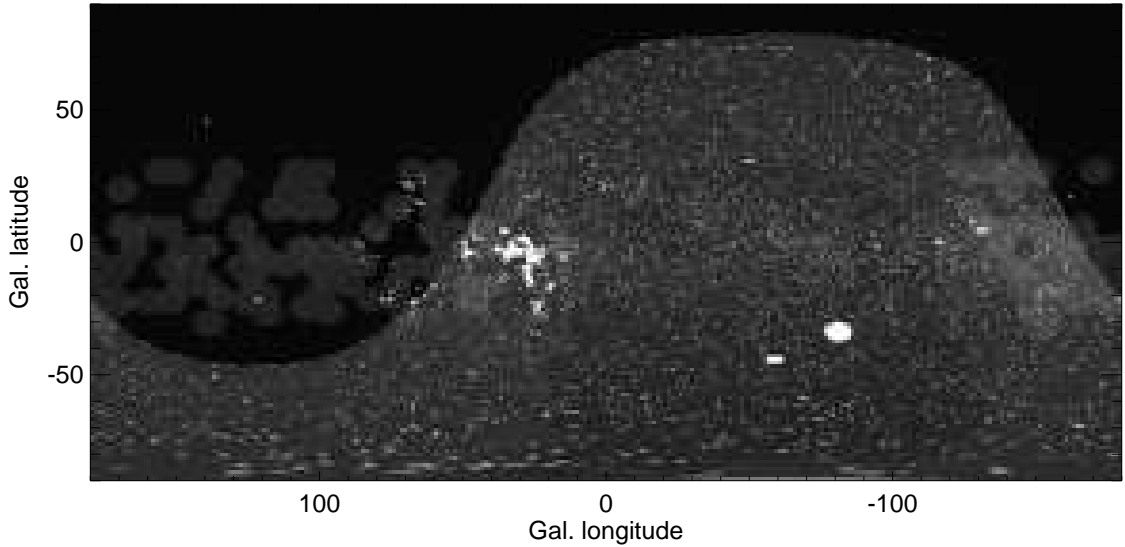}
\figcaption{Bit mask for the composite \Halpha\ map.
The Boundaries of the 3 surveys are evident, as well as the masked
point sources.  The two largest white regions are the Magellanic Clouds.
The bit mask is described in Table \ref{table_bitmask}.
\label{fig_bitmask}
}
\end{figure}

Weighting for the composite map is defined such that if SHASSA data
are available (64.0\% of the sky), they are used.  Failing SHASSA,
VTSS is used (11.7\%), and WHAM fills in the remaining 24.3\%.
The composite map is shown in Figure \ref{fig_composite} and the
bitmask in Figure \ref{fig_bitmask}.
Because the images have been zeroed to WHAM (except for 23.8\% of the
southern sky), boundaries between the high resolution images and WHAM
are as smooth as can be expected, but the sudden jump in resolution
can result in strange behavior of bright sources near the boundary.
It would be far better to have the remaining VTSS data, and
these data will be incorporated as soon as they are made public.

\begin{figure}[tb]
\epsscale{0.5}
\dfplot{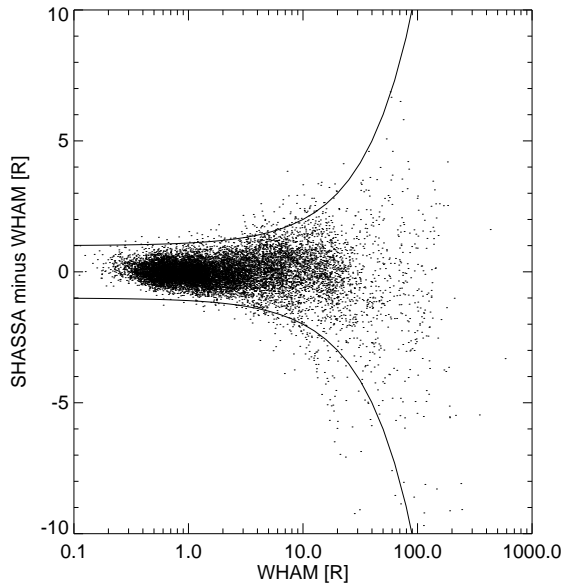}
\figcaption{Flux difference (R) of SHASSA (convolved with WHAM beam) minus
WHAM (integrated over $-80 < v_{LSR} < 80$ km/s) for every WHAM
pointing with HI\_VEL flag not set, vs. WHAM (in log R).  The lines
are $\pm(1$R + 10\%) error.  
\label{fig_calib}
}
\end{figure}

\begin{figure}[tb]
\epsscale{0.5}
\dfplot{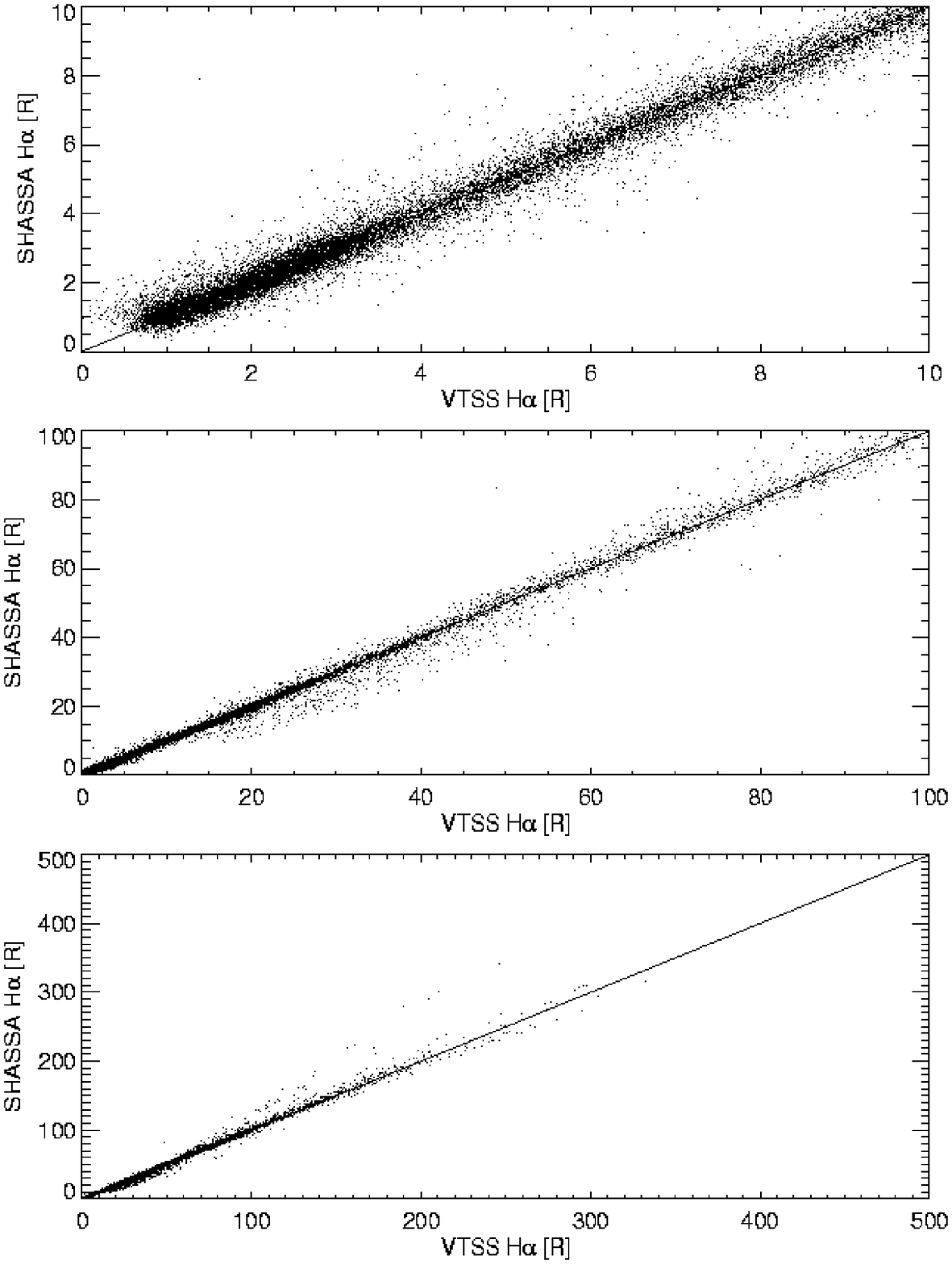}
\figcaption{Comparison of VTSS and SHASSA in the overlap region for 3 
different plot
ranges.  The WHAM zero point has been applied to both
surveys on a 1 degree scale.  For the diffuse ISM (fainter than 10R)
the agreement is excellent with a scatter of $\sim 0.3$R.  Agreement
is generally good even in the brightest regions of the sky
(see \S \ref{sec_compare}). 
\label{fig_compare}
}
\end{figure}

\subsection{Photometric calibration}
\label{sec_compare}

The overlap regions between the 3 surveys provide useful information
about relative photometric calibration.  The SHASSA zero point is
fairly good even before correction to WHAM zero point described in 
\S\ref{sec_whamzero}, so a direct comparison of the two is
straightforward.  The SHASSA minus WHAM difference is plotted in
Figure \ref{fig_calib}, where each point represents SHASSA flux
integrated over a WHAM beam minus WHAM data.  The overplotted lines
represent residuals of $\pm(1$R + 10\%) error.  
Evidently the calibration agreement is much better than 10\%, in
agreement with Gaustad \etal\ (2001). 

Because the VTSS maps were released with an arbitrary zero point, it
is more convenient to check their calibration against SHASSA directly
in the overlap region, after tying both surveys down to WHAM.  Gaustad
\etal\ found VTSS to be fainter in one overlap region by a factor of
1.25.  This same factor was applied to all VTSS data before tying down
to WHAM, and the resulting agreement between VTSS and SHASSA is good
(Figure \ref{fig_compare}).  For faint values, the two maps
agree well, with slope unity and a scatter of $\sim 0.3$R.  Brighter
than about 20R more outliers are evident, but agreement remains good
even in the brightest parts of the sky. 
The estimate of 10\% overall calibration error described in
\S\ref{sec_errors} relies on the calibration of SHASSA photometry to the 
planetary nebula photometry of Dopita \& Hua (1997), which is claimed
to be good to $\sim5\%$. 

The SHASSA data were never corrected for atmospheric extinction,
because the correction is estimated to be small by Gaustad \etal\
(2001).  They measure an extinction coefficient of
$k=0.082$ mag/airmass for \Halpha.  Since the mean airmass of
observation for SHASSA fields is always less than 1.6, this effect is
less than 5\% peak to peak.  However, the average atmospheric
absorption has been removed from the data by calibration to the 
Dopita \& Hua sources, which have been corrected to zero airmass.

\subsection{Error Map}
\label{sec_errors}

It is nearly impossible to produce an error map for the composite data
that will be meaningful for every application.  Each processing step
described above imprints another layer of covariances on the errors in
the final map, and without full knowledge of these covariances,
statistical studies (e.g. based on $\chi^2$) will be
suspect. 

Nevertheless, it is possible to estimate the uncertainty in a single
pixel on the sky, resulting from the WHAM zero point, the VTSS/SHASSA
CCD readout noise, Poisson noise from atmospheric light in the
\Halpha\ filter, and calibration errors.  Note that Poisson errors from
the Galactic \Halpha\ are negligible in the final (smoothed) maps,
being dominated by calibration uncertainty in bright regions and by
readout noise and foreground Poisson errors in faint regions.

The estimated uncertainty in each pixel is given by

\BE
\sigma = \sigma_W + N + C I(H\alpha)
\EE
where $\sigma_W$ is the error map from the WHAM data, interpolated in
the same way as the WHAM intensities; $N$ is the estimated readout
noise plus sky noise of the VTSS and SHASSA surveys (after smoothing
to a $6'$ FWHM PSF); and C is the fractional calibration uncertainty
of the VTSS and SHASSA surveys.  Within the WHAM survey area,
$\sigma_W$ is typically 0.03 R.  For pixels outside the WHAM area or
with the HI\_VEL mask bit set, $\sigma_W$ is set to 1R to 
reflect the zero-point uncertainty in the southern data
estimated by inspection.  The WHAM data products have error set to
zero for pointings contaminated by nearby stars; in these cases we
interpolate nearby error values as the correlated systematics dominate
any random measurement errors.

It is important to note that the error map does not reflect anything
about artifacts in the map, caused by saturation bleed trails, very
bright stars, etc.  For these, please refer to the bitmask, especially
the bright object bit (32) which indicates likely contamination of
these pixels by bright stars or galaxies.

\section{SUGGESTIONS FOR USE}
In this section we describe the data format of the composite \Halpha\
map, provide conversion factors from Rayleighs to EM and free-free
intensity, and discuss the effects of dust extinction on the data. 

\subsection{Projections}
The \Halpha\ map, error map, and bitmask, are provided in two formats:
a simple Cartesian Galactic longitude and latitude projection
($8640\times4320$ pixels)
and an Nside = 1024 HEALPIX\footnote{
HEALPIX is described at http://www.eso.org/science/healpix} 
(G\'{o}rski \etal\ 1998)
sphere, an equal area projection
of the sphere popular within the CMB community.  The HEALPIX
projection has 12,582,912 equal area pixels ($3.4'$ on a side) while
the Cartesian projection has square $2.5'$ pixels on the equator, with
distortions at higher Galactic latitude.  The Cartesian projection is
simple, but we encourage the use of a FITS astrometry package to
decipher the astrometric information in the file header to avoid confusion.
Both maps are presented in units of Rayleighs, where 1R = $10^6/4\pi$
photons ${\rm cm^{-2} s^{-1} sr^{-1}}$.

The Cartesian projection is the ``native'' projection of the composite
map, and the HEALPIX map is derived from it.  The Cartesian map, mask
and error map are in the files  Halpha\_map.fits, 
Halpha\_mask.fits, and
Halpha\_error.fits, respectively.
%
%

The zero-indexed fractional pixel indices (suitable for interpolation) are
\BE
x=4319.5-24l
\EE
\BE
y=2159.5+24b
\EE
for $(l,b)$ in degrees and $-180\degree < l < 180\degree$.  Note that
$l=180\degree$ is mapped to $x=-0.5$, i.e. the average of column 0 and
column 8639.  Likewise, the Galactic north and south poles are mapped
to $y=4319.5$ and $y=-0.5$ respectively.  Any interpolation code needs
to interpolate over these singularities properly. 

\subsection{Estimating Free-Free from \Halpha}
\Halpha\ and free-free are both tracers of the emission measure,
$EM=\int n_e^2dl$.  An EM of $1\emunit$ corresponds to $0.361(T/10^4\K)^{-0.9}$
R for $T<26000\K$ (Kulkarni \& Heiles). 

The emission coefficient $j_\nu$ for free-free, with electrons assumed
to interact with ions of charge $Z_i e$ and number density $n_i$ is
\BE
j_\nu = 5.44 \times 10^{-16} \frac{g_{ff}Z_i^2 n_e n_i}{T^{1/2}}
e^{-h\nu/kT} {\rm Jy~sr^{-1}~cm^{-1}}
\EE
where $g_{ff}$ is the gaunt factor for free-free.  For microwave
frequencies, a useful approximation is
\BEL{equ_gff}
g_{ff} =
\frac{3^{1/2}}{\pi}\left[\ln\frac{(2kT)^{3/2}}{\pi e^2\nu m_e^{1/2}}
 - \frac{5\gamma}{2}\right]
~~~~~\nu_p \ll \nu \ll kT/h
\EE
where $\gamma$ is the Euler constant ($\gamma\approx 0.577$) and
$\nu_p$ is the plasma frequency
(Spitzer 1978, p. 58).  For convenience, Table \ref{table_conversions}
provides coefficients for converting emission measure to free-free 
specific intensity ($I_\nu$) and brightness temperature ($T_B$).
CMB experiments usually express their results in terms of $\Delta T$,
the temperature difference relative to a 2.73K blackbody.  Brightness
temperature may be converted to this thermodynamic $\Delta T$ by
multiplying by the ``planckcorr'' factor:
\BEL{equ_planckcorr}
{\rm Planckcorr} = \frac{(e^x-1)^2}{x^2 e^x}
\EE
where $x=h\nu/k_bT_{CMB}$ and $T_{CMB}=2.73$. Table \ref{table_conversions}
displays these
factors for the nominal frequencies of by three CMB
satellites.  Note that these factors are evaluated for the frequencies
listed and are not integrated over the passbands, as should be done
for a detailed analysis.

\subsection{Ionization of Interstellar He}
The helium fraction in the interstellar medium is approximately
$Y=0.25$ by mass, or $n_{He}/n_H = 0.083$.  For calculation of the
free-free enhancement provided by the singly ionized He, the relevant
ratio is $R = N($\HeII$)/N($\HII).  We neglect doubly ionized He entirely.
The free-free emission per H atom is increased by the extra electrons
(factor of $R+1$) and by additional ions available (another factor of
$Z^2R+1$ with $Z=1$ for singly ionized He) yielding an enhancement
factor of $(R+1)^2 = 1.17$ for
$R=0.083$ (equal ionization of He and H).  However, the \Halpha\ emission
per H atom is also enhanced by a factor of $R+1$, so the ratio of
free-free to \Halpha\ is a factor of $R+1$.  In general, the value of
$R$ is less than the abundance ratio in the diffuse
ISM. 

Indeed the He ionization fraction, $\chi_{He}\equiv
n_{He^+}/n_{He(total)}$, is found to be significantly less than unity
in the diffuse WIM.  Observations of the \HeI\ $\lambda5876$
recombination line in two directions at low Galactic latitude yield
$\chi_{He} \la 0.27$ within warm, low-density regions in which the H
is primarily ionized (Reynolds \& Tufte 1995).  Radio recombination
lines provide an even tighter constraint of $\chi_{He}/\chi_{H} \la
0.13$ in the diffused ionized medium of the Galactic interior
\cite{heiles96}.  For ionization fractions this low, ionized helium is
unlikely to provide an free-free enhancement of more than a few
percent, so no correction has been made to the map.  In parts of the
sky where He is mostly ionized, a correction may be desirable.

\subsection{Combining with SFD Dust Map}
At high Galactic latitude \Halpha\ is expected to be an excellent tracer of the
warm electron emission measure, but at low latitude Galactic dust
absorbs and scatters the \Halpha\ photons so that the stated EM
derived above is too low. 

If the \Halpha\ emitting gas is uniformly mixed with dust in a
cloud with total optical depth $\tau$, the observed intensity is 
\BE
I_{\nu,obs} = \frac{I_\nu}{\tau}\int_0^\tau d\tau'e^{-\tau'}
=\frac{I_\nu}{\tau}(1-e^{-\tau})
\EE
In the limit of large $\tau$, the intensity is simply reduced by a factor
of $\tau$.

The optical depth $\tau$ is computed by multiplying the SFD dust map,
in units of $E(B-V)$ magnitudes reddening, by 2.65/1.086.  This value
is close to that obtained by evaluating the reddening law (Cardelli,
Clayton, \& Mathis 1989) for $R_V = 3.1$ at 6563\AA.  The factor of
$1.086 = 2.5\times\log_{10}e$ and converts magnitudes of extinction to
optical depth.

The derivation of the reddening law coefficient of 2.65 given above
involves careful consideration of the procedure used to calibrate the
SFD map.  The reddening map was calibrated with a sample of elliptical
galaxies on the Landolt system (see SFD, Appendix B for details).
Therefore, the CCM reddening law was multiplied by an elliptical
galaxy spectrum times the system response (atmosphere at CTIO,
telescope optics, filter, and RCA 3103A photomultiplier) defining the
Landolt system to obtain the values in SFD Table 6.  This yields the
apparently inconsistent result that $A(V)/E(B-V) = 3.315$ for Landolt
B and V, and reddening law parameter $R_V=3.1$.  This is actually
fine, because $R_V$ in this context only provides a parameterization
of the family of reddening laws, and is not literally $A(V)/E(B-V)$
for every choice of filters and source spectra.  In order to use the
SFD predictions in an internally consistent way, one must divide the
reddening law value for 6563\AA\ by the $E(B-V)$ computed by SFD,
obtaining the value of 2.65 given above.

Obviously the approximation that the dust is uniformly mixed with the
ionized gas along each line of sight will sometimes be poor.  At the
very least, the combination of the \Halpha\ map and the SFD dust map
provides a robust lower limit (dust behind) and upper limit (dust in
front of warm gas) to the actual EM, as well as a ``best guess''
(uniform mixing).  Future microwave data sets will show to what extent these
approximations are valid.

\section{SUMMARY}

We have reprocessed the VTSS and SHASSA \Halpha\ surveys to remove
bleed trails, stellar residuals, and other artifacts, and calibrated
them to the stable zero point of the WHAM survey on $1\degree$ scales.
These surveys have been combined into a well sampled $6'$ (FWHM)
full-sky map, available in two projections (Galactic Cartesian and
Galactic HEALPIX).  A bit mask summarizing the area coverage of each
survey and the processing done on each pixel is also provided, as well
as an error map.  All of these data products are available to the
public on the
web\footnote{http://skymaps.info}. 

This map is designed to have a stable zero point, minimal
contamination from stellar residuals and other artifacts, and is well
sampled (band-limited) to facilitate spherical harmonic transforms and
resampling.
These properties make the map more convenient than the original data sets as a
microwave free-free template; however, certain sacrifices have been
made.  The most obvious is that the map is somewhat lower resolution
than the maps released by the SHASSA and VTSS surveys.  In a few
places on the sky, usually near bright stars, real ISM structure has
been mistakenly removed.  Therefore, researchers interested in a
specific region of the ISM should consult the surveys directly and
take advantage of the full resolution data. 

We strongly encourage users to cite the original references describing
the three surveys, and include the acknowledgments requested by each
of them.


The Virginia Tech Spectral-Line Survey (VTSS), the Southern H-Alpha
Sky Survey Atlas (SHASSA), and the Wisconsin H-Alpha Mapper (WHAM) are
all funded by the National Science Foundation.  SHASSA observations
were obtained at Cerro Tololo Inter-American Observatory, which is
operated by the Association of Universities for Research in Astronomy,
Inc., under cooperative agreement with the National Science
Foundation.  John Gaustad, Peter McCullough, Ron Reynolds, Matt
Haffner, Bruce Draine, Ed Jenkins, David Schlegel, and Jonathan Tan
provided helpful information.  This research made use of the NASA
Astrophysics Data System (ADS) and the The IDL Astronomy User's
Library at Goddard.  Partial support was provided by NASA via grant
NAG5-6734 (Wire).  DPF is a Hubble Fellow supported by
HST-HF-00129.01-A.

\newpage

\bibliographystyle{unsrt}
\bibliography{gsrp}


\clearpage
\begin{deluxetable}{l|rrrrlr}
\footnotesize
\tablewidth{0pt}
\tablecaption{Conversion Factors from EM to Free-Free
   \label{table_conversions}
}
\tablehead{
\colhead{Experiment} &
\colhead{$\nu$} &
\colhead{$\lambda$} &
\colhead{$g_{ff}$} &
\colhead{$I_\nu$} &
\colhead{$T_{B}$} &
\colhead{$T_{thermo}$}
\\
\colhead{} &
\colhead{GHz} &
\colhead{mm} &
\colhead{} &
\colhead{$\JypSr$} &
\colhead{$\microK$} &
\colhead{$\microK$}
\\
\colhead{} &
\colhead{(2)} &
\colhead{(3)} &
\colhead{(4)} &
\colhead{(5)} &
\colhead{(6)} &
\colhead{(7)}
}
\startdata
COBE/DMR  &   31.5 &   9.52 &  4.058 &  68.65 &   2.255 &   2.313 \\
          &   53.0 &   5.66 &  3.771 &  63.79 &   0.740 &   0.795 \\
          &   90.0 &   3.33 &  3.479 &  58.84 &   0.237 &   0.290 \\ \hline
MAP       &   22.0 &  13.64 &  4.256 &  72.00 &   4.849 &   4.909 \\
          &   30.0 &  10.00 &  4.085 &  69.10 &   2.503 &   2.561 \\
          &   40.0 &   7.50 &  3.926 &  66.42 &   1.353 &   1.410 \\
          &   60.0 &   5.00 &  3.703 &  62.63 &   0.567 &   0.622 \\
          &   90.0 &   3.33 &  3.479 &  58.84 &   0.237 &   0.290 \\ \hline
Planck    &   30.0 &  10.00 &  4.085 &  69.10 &   2.503 &   2.561 \\
          &   44.0 &   6.82 &  3.874 &  65.53 &   1.103 &   1.159 \\
          &   70.0 &   4.29 &  3.618 &  61.19 &   0.407 &   0.461 \\
          &  100.0 &   3.00 &  3.421 &  57.86 &   0.189 &   0.242 \\
          &  143.0 &   2.10 &  3.224 &  54.51 &   0.0869 &   0.143 \\
          &  217.0 &   1.38 &  2.994 &  50.60 &   0.0350 &   0.104 \\
          &  353.0 &   0.85 &  2.726 &  46.04 &   0.0120 &   0.154 \\
          &  545.0 &   0.55 &  2.486 &  41.96 &   0.00460 &   0.727 \\
          &  857.0 &   0.35 &  2.237 &  37.69 &   0.00167 &  25.746 \\ \hline
\enddata
\tablecomments{
Col. (2): Nominal central frequency, in GHz.
Col. (3): Corresponding wavelength, in mm. 
Col. (4): Gaunt factor for free-free, given in eq.(\ref{equ_gff}).
Col. (5): Specific intensity, $I_\nu$, corresponding to EM=1.
Col. (6): The brightness temperature ($\mu$K) corresponding to EM=1.
Col. (7): Brightness temperature multiplied by Planckcorr (eq.
\ref{equ_planckcorr}) to convert to
thermodynamic $\Delta T$, assuming $T_{CMB}=2.73$K.}
\end{deluxetable}
\clearpage
\begin{deluxetable}{l|l|r|l}
\footnotesize
\tablewidth{0pt}
\tablecaption{Bitmask Values
   \label{table_bitmask}
}
\tablehead{
\colhead{Bit} &
\colhead{Name} &
\colhead{\% Sky} &
\colhead{Comments}
}
\startdata
     0 & WHAM     & 76.23 & WHAM data used (Wisconsin)                       \\
     1 & VTSS     & 17.35 & Virginia Tech Spectral line Survey               \\
     2 & SHASSA   & 63.99 & Southern Halpha Sky Survey Atlas                 \\
     3 & STAR     &  6.94 & Star removed                                     \\
     4 & SATUR    &  2.58 & Saturated pixel nearby in continuum exposure     \\
     5 & BRT\_OBJ &  1.90 & Bright star/galaxy - measurements unreliable     \\
     6 & BIG\_OBJ &  0.14 &Position near LMC, SMC, or M31; no sources removed\\
     7 & HI\_VEL  &  0.53 & High Velocity in WHAM data             \\
\enddata
\tablecomments{
The bitmask contains important information about survey coverage (bits
0-2), artifact removal (3,4) or lack thereof (6) and reliability (5).
The BRT\_OBJ and BIG\_OBJ masks in particular must be used for any
full-sky statistical study.  Regions with HI\_VEL set may have WHAM 
zero-point errors of greater than 10\% and 0.2R.  An image of the
bitmask is shown in Figure \ref{fig_bitmask}.
}
\end{deluxetable}
\clearpage

\end{document}